\documentclass[12pt]{article}
\usepackage{graphicx}

\begin{document}

\newcommand{\tableheadseprule}{\hline}

\lefthyphenmin=2
\righthyphenmin=2

\title{\bf Learning intrinsic excitability in medium spiny neurons}

\author{Gabriele Scheler\\
Carl Correns Foundation for Mathematical Biology, \\
Mountain View, Ca, USA\\
E-mail: gscheler@gmail.com
}

\date{}

\maketitle

\begin{abstract}
We present an unsupervised, local activation-dependent
learning rule for intrinsic plasticity (IP) which affects the 
composition of ion channel
conductances for single neurons in a use-dependent way.

We use a single-compartment conductance-based model for medium spiny
striatal neurons in order to show the effects of parameterization of 
individual ion channels on the neuronal activation function.
We show that parameter changes within the physiological ranges  
are sufficient to create an ensemble of neurons 
with significantly different activation functions.
We emphasize that the effects of 
intrinsic neuronal variability on spiking behavior require a distributed 
mode of synaptic input and can be eliminated by strongly correlated input.

We show how variability and adaptivity in ion channel conductances 
can be utilized to store patterns without an additional
contribution by synaptic plasticity (SP). 
The adaptation of the spike response may result in either "positive" 
or "negative" pattern learning.
However, read-out of stored information depends
on a distributed pattern of synaptic activity to let intrinsic variability
determine spike response. We briefly discuss the implications
of this conditional memory on learning and addiction.
\end{abstract}

%============================================================
\section{Introduction}

A role for modification of activation functions, or intrinsic plasticity (IP), 
for behavioral learning has been demonstrated for a number of 
systems \cite{ZhangLinden2003,Mahon2012,Rosenkranz2011}.
For instance, in rabbit eyeblink conditioning,
when ion channels related to afterhyperpolarization are being suppressed by a
learning event, they can become permanently suppressed.
This has been shown for pyramidal cells of hippocampal areas CA1 and CA3,
and for cerebellar Purkinje cells \cite{SchreursBGetal98,Coop2010}.
In some cases, these
changes are permanent and still present after 30 days
\cite{MoyerJRetal96,ThompsonLTetal96}, in other cases, 
intrinsic changes disappear after 3-7 days, while the behavioral
memory remains intact, raising questions about the long-term component of
intrinsic plasticity in these systems.
There are at the present time conflicting ideas on the significance
of IP compared to synaptic plasticity
\cite{ZhangLinden2003,Azdad2009}, and the range of functions that IP may have in
adaptivity \cite{DestexheMarder2004,Xu2005,Baroni2010,Naude2012}.

A few computational models have been proposed that show how modification
in activation functions can be achieved with 
ion channel based models of realistic single neurons.
Marder and colleagues have developed an approach, where they sample 
a very large parameter space for conductances of ion channels,
exploring nonlinearities in the relation between conductances and neural 
spiking behavior 
\cite{Marder96,GoldmanMSetal2001,PrinzAetal2004}.
The motivation for this research are observations about neuromodulation 
and intrinsic plasticity in specific neurons of an invertebrate 
ganglion (e.g., \cite{LeMassonGetal93}).
They have noted that large variations in some parameters may have little
effect on neuronal behavior, while comparatively small variations in certain
regions in parameter space may change response properties significantly.
They also suggest that neuromodulation may provide an efficient means of
targeting regions in parameter space with significant effects on response
properties \cite{GoldmanMSetal2001}.

A study by \cite{StemmlerKoch99} assumed 
the goal of modification of activation functions is to  
achieve an optimal distribution of firing rates for a population of neurons. 
The idea was that by tuning each neuron 
to a different band of the frequency spectrum, the full bandwidth of 
frequencies could be employed for information transfer. This goal was 
achieved by adjusting $Na^+$, $K^+$, and $Ca^{++}$ channels for a generically defined 
neuron until a desired frequency was stably reached.

We present a different approach, where the modification of activation functions 
reflects the history of exposure to stimuli for a specific neuron.
Similarly, \cite{DaoudalDebanne2003,Disterhoft2006} suggested that synaptic LTP/LTD 
and linear regulations of intrinsic excitability could operate in a 
synergistic fashion.
However, in our approach, different types of synaptic stimulation result 
in state changes for the neuronal unit, influencing its capacity for 
read-out of stored intrinsic properties. Thus, intrinsic plasticity is
conceptualized as fundamentally different from synaptic plasticity which does not 
encompass such a state change.
The learning rule that we derive as the basis for 
adjustment concerns one-dimensional upregulation or down-regulation of 
excitability in the "read-out" state of the neuron, and affecting only this 
state.
This rule uses neural activation, significantly determined by intracellular 
calcium for the learning parameter, which can
be shown to be biologically well-motivated (cf. also \cite{LeMassonGetal93}).

%============================================================
\section{Materials and Methods}
\subsection{Striatal medium spiny neuron}
The membrane voltage $V_m$ is modeled as
\[ \dot{V_m} = -{1 \over C} [\sum_i{I_i} - I_{syn}]\]
The individual currents are modeled by conductances, state variables and 
the reversal potential:
\begin{equation}
 I_i = \bar{g_i}(V_m) * m^{p_i}*h_i^{q_i}*(V_m - E^{rev}_i) 
\label{eq:basic}
\end{equation}
The dynamics are defined using state variables for activation ($m$) and
inactivation ($h$). The types of equations used for the dynamics are:

\begin{enumerate}
\item exponential: $f(V_m) = \lambda \exp({{V_m - V_i}\over -V_c})$
\item logistic: $f(V_m) = {\lambda \over 
{1+ \exp({{V_m - V_i}\over -V_c})}}$
\item linexp: $f(V_m) = {{\lambda (V_m - V_i)} \over 
        {1+ \exp({{V_m - V_i}\over -V_c})}}$
\end{enumerate}

The state variables can be defined indirectly using 
\[ \dot{m} = (1-m)\alpha - m \beta \] 
and 
\[ \dot{h} = (1-h)\alpha - h \beta \]
and one of the equations (1-3) with different values for $\lambda$ 
($\lambda_{\alpha}, \lambda_{\beta}$), $V_i$ ($V_i^{\alpha}, V_i^{\beta})$ 
and $V_c$ ($V_c^\alpha$, $V_c^\beta$).
We use this method for the 
ion channels in Table~\ref{wangbuzsaki}.

The state variables can also be directly defined 
(cf. \cite{GoldmanMSetal2001}):
\[ \dot{m} = {{m_\infty - m} \over \tau_m}\]
\[ \dot{h} = {{h_\infty - h} \over \tau_h}\] 
The parameters used are $m_\infty=m0, h_\infty=h0, \tau_m$ and $\tau_h$ as 
in Table~\ref{Mahon}. Again, we use one of the equations (1-3)
with the $\lambda$ parameters ($\lambda_{m0}$ and $\lambda_{h0}$) set to 1.
(These representations are mathematically equivalent and related by
$ m_\infty  = {\alpha_m \over {\alpha_m + \beta_m}}$, 
$ \tau_\infty  = {1\over {\alpha_m + \beta_m}} $.)

Standard parameter values for the modeling of ion channels ('naive 
state') were compared with several publications.
Parameter values for $I_{K}$, $I_{Na}$ and $I_{leak}$
were adapted from \cite{WangBuzsaki96}, 
for L-type calcium channels ($I_{CaL}$) from \cite{BargasJetal94} and 
\cite{TsuboYetal2004},
see Table~\ref{wangbuzsaki}.

%-------------------
% TABLE 1
%-------------------

Parameters for slow A-type K channels ($I_{As}$) were adapted 
from \cite{GabelNisenbaum98,NisenbaumESetal98}, 
for fast A-type K channels ($I_{Af}$) from
\cite{SurmeierDJetal89}, 
for inward rectifying K channels ($I_{Kir}$) from \cite{NisenbaumWilson95},
and the resulting parameter tables were compared with
\cite{MahonSetal2000} and \cite{GruberAJetal2003}, see Table~\ref{Mahon}. 

%-------------------
% TABLE 2
%-------------------

\subsection{Variability} 
\label{variability}
The maximum conductance of different ion channels can be expressed by a 
scaling factor in the membrane potential equations as in 
Eq.~\ref{eq:mu-factorsI}
(for synaptic currents) or Eq.~\ref{eq:mu-factorsg} (for synaptic 
conductances), cf. \cite{GruberAJetal2003}.
\begin{equation}
{\dot V_m} = -{ 1 \over C} [\mu_1 I_1  + \mu_2 I_2 \ldots + \mu_i I_i - I_{syn}] 
\label{eq:mu-factorsI}
\end{equation}

\begin{equation}
{\dot V_m} = -{1 \over C} [\mu_{1} I_{1} + \ldots +  g_s (V_m-V_0)]
\label{eq:mu-factorsg}
\end{equation}

Both NM-independent and NM-dependent modifications may coexist in a neuron,
as expressed in Eq.~\ref{eq:mu-kappa-factors} 
([NM] stands for the level of synaptic availability of a neuromodulator 
NM).

\begin{equation}
{\dot V_m} = - {1 \over C} [(\mu_1 I_1 + [NM] \kappa_1 I_1) + (\mu_2 I_2 + [NM] \kappa_2 I_2) \ldots]
\label{eq:mu-kappa-factors}
\end{equation}
In this paper, for 
simplicity, we shall 
refer to (Eq.~\ref{eq:mu-factorsI}) as the generic format for 
intrinsic adaptation, with the understanding that $\mu$ is replaceable by 
$[NM]\kappa$.

Physiological ranges for $\mu$ can be estimated by various 
means. 
There are measurements for 
variability in electrophysiologically defined membrane behavior (current
threshold, spike response to current pulses etc. 
\cite{OnnSPetal2003,MahonSetal2000b}) that 
are typically expressed as standard errors (e.g. 16-20\% for current 
threshold, \cite{OnnSPetal2003}). There are 
also attempts at classifying MSN cells into different 'types' based on 
their electrophysiological profile \cite{WickensWilson98,OnnSPetal2000}.
Modeling shows that 
variability of ion channel conductances with a range of $\pm 40\% $
matches 
measures of electrophysiological variability and reproduces the ranges 
for MSN types (data not shown). 
Interestingly, direct measurements for dopamine D1 receptor-mediated 
changes on ion channel conductances are approximately in the same ranges 
($\pm 30-40\%$, \cite{GruberAJetal2003}).
Our discussion is thus based on an 
estimate of $\mu$ ranging from $0.6-1.4$ for each channel.

\subsection{Defining synaptic input}
Synaptic input is defined by overlays of the EPSPs generated by $N$ individual
Poisson-distributed spike trains with mean interspike interval $\tau_{syn}$.
Each EPSP is modeled as a spike with peak amplitude $I_0 = 1.2\mu A/cm^2$ and
exponential
decay with $\tau=2.5ms$ similar to \cite{TsuboYetal2004,SachdevRNetal2004}.
IPSPs are modeled in a similar way with $I_0=-0.4\mu A/cm^2$.
This corresponds to 0.5nA (-0.2nA) as peak current (with $1nA=2.3 \mu A/cm^2$).
Synaptic conductances are calculated by
$g_{syn} = I_{syn}/(V_m - V_0)$ with $V_0$ set to 0mV. 
We have tuned the model to 
$g_{syn}$= $0.0035mS/cm^2$ for a first spike for the naive or standard neuron 
(all $\mu=1$). 
At -40mV, this is $0.0035mS/cm^2 * (-40mV) = -1.4 \mu A/cm^2$ or $0.6nA$,
which corresponds to the experimentally measured average value 
for the rheobase in \cite{OnnSPetal2003}.
We may increase the correlation in the input by using a percentage $W$ of 
neurons which fire at the same time.
Higher values for $W$ increase the 
amplitude of the fluctuations of the input (cf. \cite{BenucciAetal2004}).

\subsection{Implementation}
%The simulator has been implemented in Matlab and executed
%on an Apple 1GHz G4 notebook.
%The entire code is interpreted and no specific code optimizations
%have been applied. For numerical integration, the solver ode45 was used.
%Simulation of one neuron for 1s took approximately 68s CPU time. It is 
%expected that for compiled and optimized code the simulation speed
The simulator has been implemented in Matlab.
The entire code is interpreted and no specific code optimizations
have been applied. For numerical integration, the solver ode45 was used.

%============================================================
\section{Results}
\subsection{Intrinsic Variability}

\begin{figure}[htp]
\begin{center}
\includegraphics[width=0.95\textwidth]{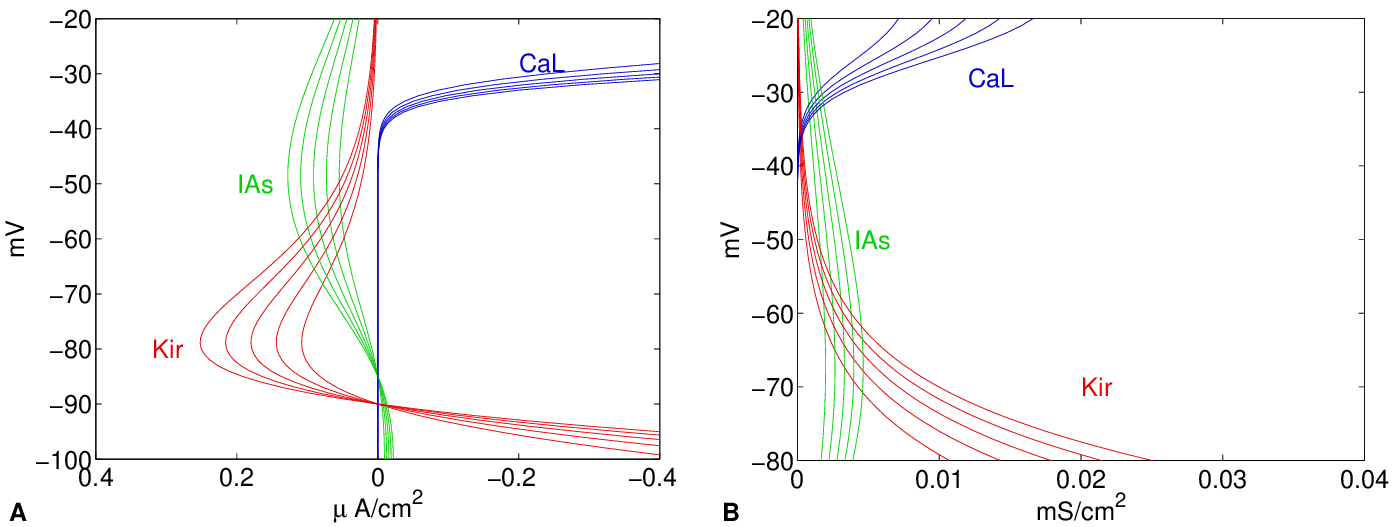}
\end{center}
\caption{{\bf Variability of ion channel density:} Variable factors ($\mu$ = \{0.6 ... 1.4\}) for the slowly
inactivating $K^+$-channel (Kv1.2, $I_{As}$), the L-type calcium channel
($I_{CaL}$), and the inward rectifying K+ channel ($I_{Kir}$) are shown at 
different membrane voltages $V_m$ (A) in an I-V plot, (B) 
as variability in conductance.}
\label{KAs-conductance}
\end{figure}

\begin{figure}[htp]
\begin{center}
\includegraphics[width=0.95\textwidth]{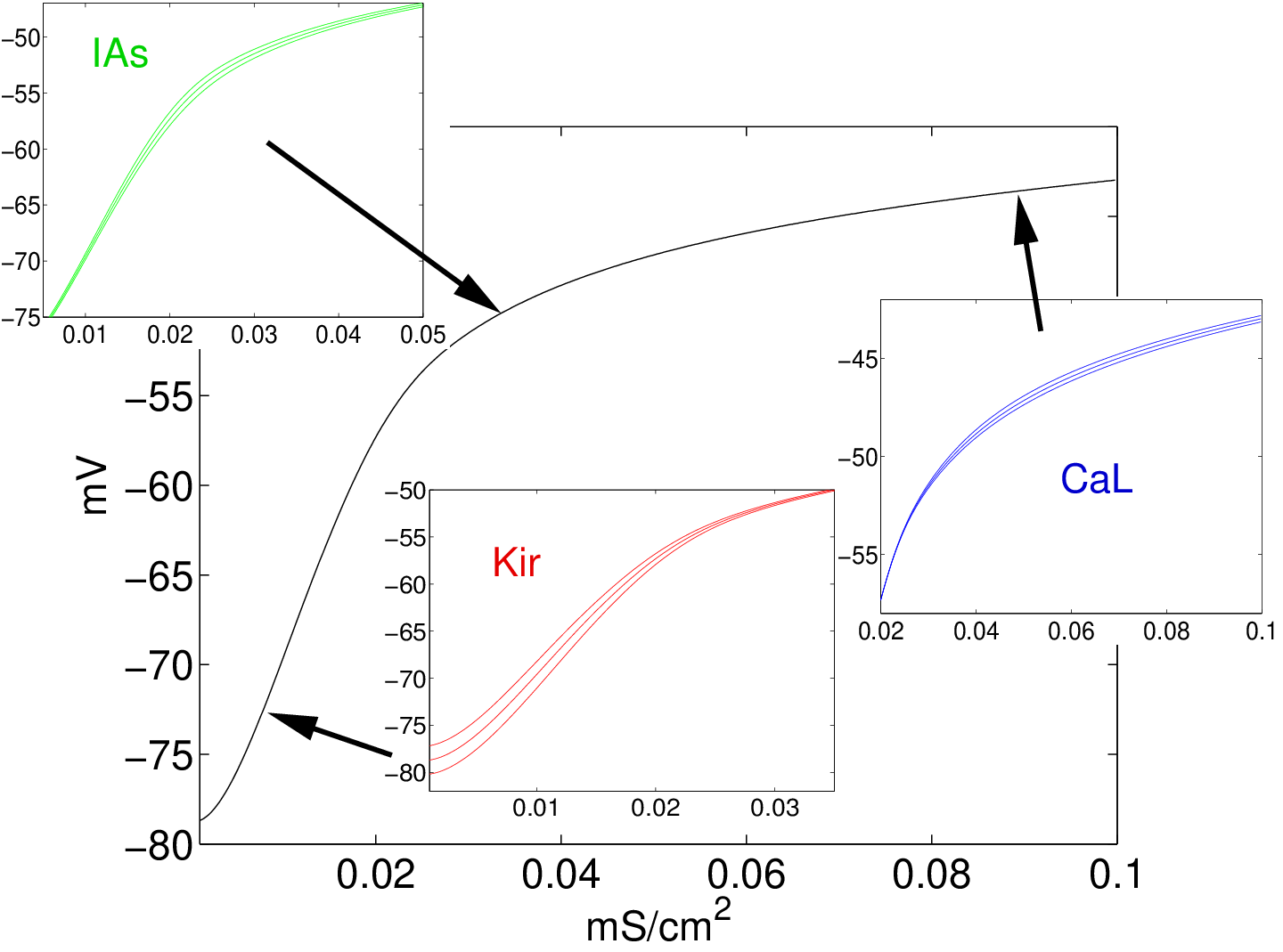}
\end{center}
\caption{{\bf Variability of activation function:} Variable factors ($\mu$ = \{0.6 ... 1.4\}) for 
$I_{As}$, $I_{CaL}$, and $I_{Kir}$  
as components of the activation function ($g_s$ vs. $V_m$).
The activation function is defined as the membrane voltage response 
for different injected (synaptic) conductances ($g_s$), and computed 
by solving Eq~\ref{eq:mu-factorsg} for the membrane voltage $V_m$.}
\label{actfun}
\end{figure}

We explore the impact of small variations in ion channel
conductances on the shape of the activation function.
As an example, we show the current and conductance changes for a slowly
inactivating A-type K+ channel (Kv1.2, $I_{As}$), L-type calcium
channel ($I_{CaL}$) and inward rectifying K+ channel ($I_{Kir}$)
at different
membrane potentials modulated by a scaling factor 
$\mu = \{$0.6, 0.8, 1.0, 1.2, 1.4$\}$
(Fig.~\ref{KAs-conductance}, Fig.~\ref{actfun}). 
Regulation of the voltage-dependence
\cite{MisonouHetal2004} and even of the inactivation dynamics of 
an ion channel 
\cite{HayashidaIshida2004} has also been shown, but these effects 
are not further discussed here. 

%----------------
% Fig 1,2
%----------------
%\caption{Variable factors ($\mu$ = \{0.6 ... 1.4\}) for the slowly
%inactivating $K^+$-channel (Kv1.2, $I_{As}$), the L-type calcium channel
%($I_{CaL}$), and the inward rectifying K+ channel ($I_{Kir}$) are shown at 
%different membrane voltages $V_m$ (top) in an I-V plot, (bottom) 
%as variability in conductance.}
%\caption{Variable factors ($\mu$ = \{0.6 ... 1.4\}) for 
%$I_{As}$, $I_{CaL}$, and $I_{Kir}$  
%as components of the activation function ($g_s$ vs. $V_m$).
%The activation function is defined as the membrane voltage response 
%for different injected (synaptic) conductances ($g_s$), and computed 
%by solving Eq\ref{eq:mu-factorsg} for the membrane voltage $V_m$.}

We can see that there are critical
voltage ranges (a\-round -50mV, around -80mV and starting at -40mV), where the conductance and the current are
highest, and where scaling has a significant effect, while scaling 
has small or no effect in other voltage ranges. (The Na+ current
has been disabled for this example to prevent the neuron from firing).

\begin{figure}[htp]
\begin{center}
\includegraphics[width=0.95\textwidth]{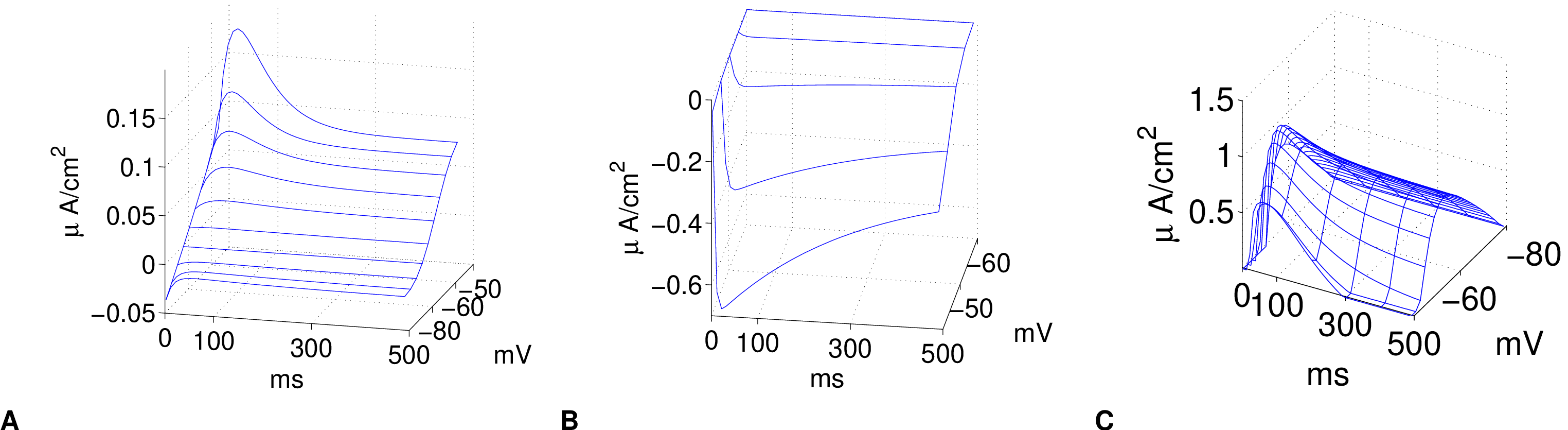}
\end{center}
\caption{{\bf Activation-inactivation (temporal) dynamics:} (A) 
dynamics for the slow A channel $I_{As}$ 
(B) 
the L-type Ca channel $I_{CaL}$, and (C) for the set of ion channels 
used in the standard MSN model. We see a rise time due to $I_{As}$ and 
overlapping inactivation dynamics in the -55 to -40 mV range.}
\label{temporal}
\end{figure}

In Fig.~\ref{temporal},  we show 
the current over time---to graphically display the slow dynamics 
of the $I_{As}$ and $I_{CaL}$ channel. 
Since we do not change the 
activation-inactivation dynamics of any channel in our model, we show currents 
only for $\mu_{As}$, $\mu_{CaL} = 1$.

We can see that $I_{As}$ activates 
moderately fast (20ms), while it inactivates with a half-time of 
about 300ms, depending on the voltage. For $I_{CaL}$, activation is almost 
instantaneous,
but inactivation is $>$  500ms. 

%----------------
% Fig 3
%----------------
%\caption{Activation-inactivation (temporal) dynamics (A) 
%for the slow A channel $I_{As}$ 
%(B) 
%the L-type Ca channel $I_{CaL}$, and (C) for the set of ion channels 
%used in the standard MSN model. We see a rise time due to $I_{As}$ and 
%overlapping inactivation dynamics in the -55 to -40 mV range.}
%\label{temporal}

The activation function for the 
MSN model shows a time-dependence only in the high-voltage range (at 
or above -55mV), whereas the components in the lower voltage ranges are 
not time-dependent.

Mathematically, we can consider the individual channels
as a set of basis functions that allow function approximation for the 
activation function. Each particular adjustment of an activation 
function can be considered learning of a filter which is suited to a 
specific processing task. The activation-inactivation dynamics would provide a
similar set of basis functions for the temporal domain. Of course, it 
is interesting to note which particular basis functions exist, and also 
how the temporal dimension is tied in with specific 
voltage-dependences. For instance, 
the slowly inactivating potassium channel $I_{As}$ provides a skewed mirror 
image of the function of calcium-gated Sk/BK channels, which are 
responsible for afterhyperpolarization, making different variants of 
frequency filters possible. 
On this basis, a mapping of ion channel components 
and their density or distributions in different types of neurons could 
provide an interesting perspective on direct interactions for neurons from 
different tissue types or brain areas, as well as e.g. between cholinergic 
interneurons and MSNs within striatum.

\begin{figure}[htp]
\begin{center}
\includegraphics[width=0.95\textwidth]{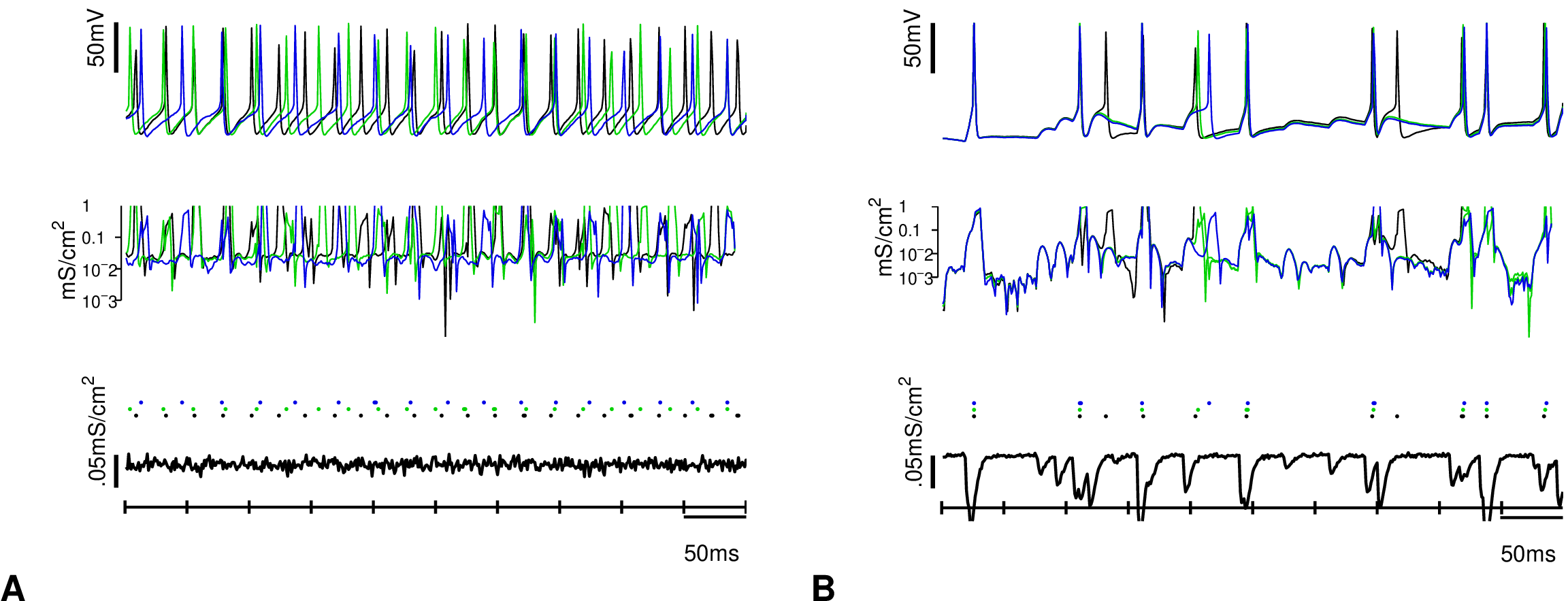}
\end{center}
\caption{{\bf Input correlation-dependent read-out of intrinsic memory:} Response to inputs generated from $N=80$ neurons with independent
Poisson processes using different
correlations parameters $W=0.2, 0.9$ (A, B). 
Three slightly different neurons with $\mu_{As} = 1.1, 1.3, 1.5$ are shown 
under BOTH conditions. 
(A) Response variability and different firing rates for each neuron 
(here: 20, 26, 40Hz) occur with 
distributed (low correlation) input.
(B) Highly correlated input produces reliable spiking and by implication a 
single firing rate (20Hz).
The upper panel shows the membrane voltage, the middle panel shows the membrane 
conductances, and the lower panel shows the synaptic input as conductance.}
\label{currentgsyn}
\end{figure}

To further explore the influence of variability of the activation 
function, we apply realistic synaptic input with different amounts of
correlation to individual MS neurons (see Fig.~\ref{currentgsyn}).

%------------------------
% Fig 6
%------------------------
%\caption{Response to inputs generated from $N=80$ neurons with independent
%Poisson processes using different
%correlations parameters $W=0.2, 0.9$ (A,B). 
%Three slightly different neurons with $\mu_{As} = 1.1, 1.3, 1.5$ are shown 
%under BOTH conditions. 
%(A) Response variability and different firing rates for each neuron 
%(here: 20,26,40 Hz) occur with 
%distributed (low correlation) input.
%(B) Highly correlated input produces reliable spiking and by implication a 
%single firing rate (20Hz).
%The upper panel shows the membrane voltage, the middle panel shows the membrane 
%conductances, and the lower panel shows the synaptic input as conductance.}
%\label{currentgsyn}

This shows us that small adjustments in the contribution of a specific ion 
channel can result in significantly different spiking 
behavior even for identical synaptic input. 
This occurs when the input is distributed, i.e. has low correlation. In this 
case, the neurons spike independently of each other and with different 
frequencies.
We can eliminate this effect by increasing the correlation of the input.
Because of the slow activation/inactivation dynamics of the $I_{As}$ channel,
(latency of $\approx$ 20ms)
only low correlated input activates these 
channels ('neuronal integrator mode'), but
highly correlated inputs do not activate these channels, driving the membrane 
to spiking quickly ('coincidence detector mode').
Therefore correlated input can produce reliable spiking behavior for model 
neurons which differ in the relative contribution of the slow $I_{As}$ channel. 
Distributed input, in contrast, activates slower ion channels, 
and can produce different tonic firing rates, here according to 
the contribution of the $I_{As}$ channels, as long as strong synaptic 
input keeps the neuron in the relevant voltage range ('persistent activity'). 

Similarly, the differential contribution of other channels 
(high-voltage gated L-type Ca-channels, hy\-per\-po\-la\-riz\-ation-activated 
GIRK channels or calcium-dependent Sk/Bk channels)
will affect neuronal behavior, when the conditions for a prominent 
influence of these channels are met.

We are modeling a state of MS neurons that exhibits regular tonic firing.
In experimental studies, \cite{WilsonKawaguchi96}
showed that MS neurons, 
similar to cortical neurons, exhibit upstate-downstate behavior, reminiscent 
of slow-wave sleep, under certain forms of anesthesia (ketamine, barbiturate).
However, under neurolept-analgesia (fent\-anyl-halo\-peri\-dol),
\cite{MahonSetal2003} showed that MS neurons can show driven activity, when 
cortical input is highly synchronized, and exhibit a state characterized 
by fluctuating synaptic inputs without rhythmic activity (i.e. 
without upstates/downstates), when cortical input is desynchronized. 
The regular, tonic spiking in 
this state is very low, much less than in the waking animal,
which may be related to the dopamine block by haloperidol.
%the effects of which would be strongest in striatum, rather than cortex. 
This makes a waking state of MS neurons characterized 
by regular tonic spiking at different firing rates probable.

In the following, we show how intrinsic  
excitability adaptation can lead to different recalled firing rates under
appropriate synaptic stimulation. 
The model could thus reflect learning that is recalled or read out 
during MSN states under desynchronized cortical input - in contrast 
to highly synchronized input, which would homogenize the response of 
the coincidence detecting neurons and favor reliable transfer of spikes.

\subsection{Induction and Maintenance of Plasticity}
The general idea for learning intrinsic plasticity is to use a learning 
parameter $h$ for each individual update of the conductance scaling 
factor $\mu$.
The {\it direction} of learning ($h>0$ or $h<0$) is determined from the 
neural activation ($A_n$) for each individual neuron.
Neural activation is largely determined by intracellular calcium 
but here we estimate 
the neural activation $A_n$ from the spike rate of the neuron, measured 
over 1s of simulated behavior (see Section~\ref{biolearn} for a discussion).

We define a bidirectional learning rule dependent on an initial firing 
rate $\theta$:
excitability is increased by a step function $h$ (with  
stepsize $\sigma$)
when $A_n$ is greater than $\theta$, excitability is decreased
when $A_n$ is lower than $\theta$
("positive learning").
This means, when the actual neural activation is higher than the initial 
firing rate, membrane adaptations aim to move the neuron to a higher 
excitability in order to create a positive memory trace of a period of 
high activation (which can then be replicated under distributed 
synaptic stimulation). The same mechanism applies to lower the excitability 
of a neuron.
\begin{equation}
\begin{array}{ll}
 \Delta \mu = h(A_n) = \sigma & \mbox{if}\  A_n > \theta \\ 
 \Delta \mu = h(A_n) = -\sigma & \mbox{if}\  A_n < \theta\\
\end{array}
\label{eq:An-learning}
\end{equation}

This rule can also be implemented by individual increases in excitability
after each action potential, and decreases of excitability for periods of
time without action potentials. 
Initial experiments \cite{MahonSetal2003,ZhangWetal2004} 
indeed show such adaptation of intrinsic excitability after individual 
spikes.

The function $h$ can be applied to a single ion channel, such as 
$I_{As}$, but also to a number of ion channels in parallel: 
e.g. to mimic dopamine D1 receptor activation, $h$ may be applied to  
$\mu_{As}$ (upregulated with high $A_n$), $\mu_{Na}$ 
(downregulated with high $A_n$), and $\mu_{CaL}$ (downregulated with 
high $A_n$). 

\subsection{Pattern Learning}
\begin{figure}[htp]
\begin{center}
\includegraphics[width=0.95\textwidth]{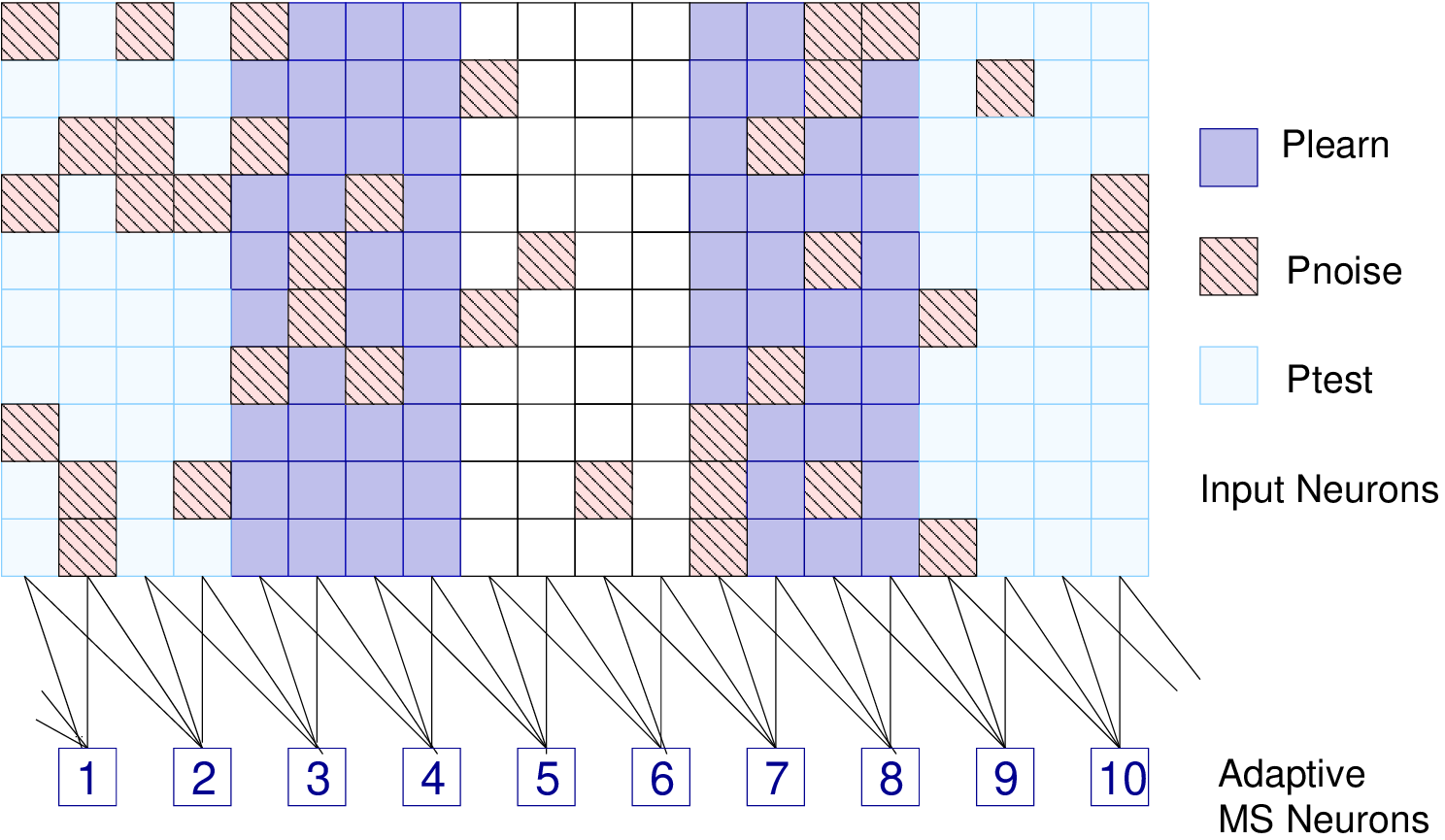}
\end{center}
\caption{{\bf Pattern learning:} 200 input neurons (arranged as 20X10), 10 
learning neurons, and definitions for 3 patterns. Only 37 of 80 data points 
for $P_{noise}$ are shown.}
\label{toplo}
\end{figure}

We can show the effect of this learning 
rule on pattern learning.
We generate synaptic inputs from a grid of 200 input neurons for a single 
layer of 10 MSNs. On this grid we project two stripes of width 4 
as a simple input pattern $P_{learn}$  by adjusting 
the mean interspike interval (ISI) for the corresponding 
input neurons to a higher value ($ISI = 350$ ms for {\it on} 
vs. $750$ ms for {\it off} neurons, see Fig.~\ref{toplo}).

%----------------
% Fig toplo
%----------------
%\caption{Pattern learning: 200 input neurons (arranged as 20x10), 10 
%learning neurons, and definitions for 3 patterns. Only 37 of 80 data points 
%for $P_{noise}$ are shown.}
%\label{toplo}

We apply the learning rule to each of the currents
$I_{Na}$, $I_{CaL}$ and $I_{As}$. This mimics changes in dopamine D1 receptor 
sensitivity, which targets these ion channels. Adaptation can be 
weaker or stronger, depending on learning time
(e.g., $\sigma$=0.01, $t$=20s (20 steps) (weak), $t$=40s (40 steps) (strong)).
After a number of steps, we achieve a distribution of $\mu$-values 
that reflects the strength of the input (Table~\ref{table3}A).

\begin{figure}[htp]
\begin{center}
\includegraphics[width=0.95\textwidth]{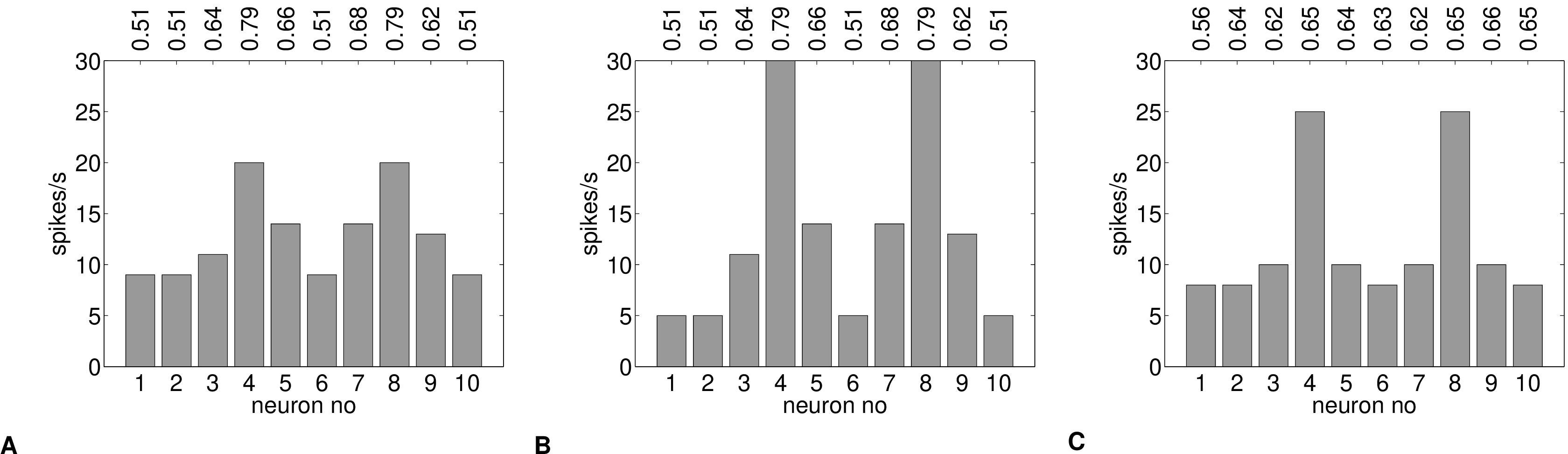} 
\end{center}
\caption{{\bf Positive pattern learning:} spike frequency histograms 
for 10 adaptive neurons ($\theta$=11.5Hz)
(A) response of naive neurons to $P_{learn}$ (B) 
response of $P_{learn}$-adapted neurons to $P_{learn}$ 
(C) $P_{learn}$-adapted neurons tested with $P_{noise}$. 
Average synaptic input ($nA/cm^2$) for each neuron is shown on top. 
Responses in (A) 
and (B) to the same input $P_{learn}$ are different, a pattern similar 
to $P_{learn}$ emerges in response to uniform (noise) pattern input in (C). 
}
\label{positive}
\end{figure}

In Fig.~\ref{positive}, we obtain spike frequency histograms from
the set of MS neurons under different conditions.
Fig.~\ref{positive}A shows the naive response to the input pattern
$P_{learn}$ - high activation in two medial areas.
After adaptation, this response is increased (Fig.~\ref{positive}B).
When we apply
a test input of a random noise pattern $P_{noise}$, 
we see that the learned pattern is still reflected in the spike histogram 
(Fig.~\ref{positive}C).
For positive learning, this process is theoretically unbounded, and 
only limited by the stepsize and the adaptation time. A saturation state 
could be defined to prevent unbounded learning, which would also allow 
to perform capacity calculations.

%----------------
% positive
%----------------
%\caption{Positive pattern learning: spike frequency histograms 
%for 10 adaptive neurons ($\theta$=11.5Hz)
%(A) response of naive neurons to $P_{learn}$ (B) 
%response of $P_{learn}$-adapted neurons to $P_{learn}$ 
%(C) $P_{learn}$-adapted neurons tested with $P_{noise}$. 
%Average synaptic input ($nA/cm^2$) for each neuron is shown on top. 
%Responses in (A) 
%and (B) to the same input $P_{learn}$ are different, a pattern similar 
%to $P_{learn}$ emerges in response to uniform (noise) pattern input in (C). 
%}
%\label{positive}

We should note that applying just one pattern continuously results in
a very simple learning trajectory: each update results in a step 
change in the relevant ion channel currents.
However, we also show that the effects of stepwise adaptation of individual 
ion channels do not necessarily lead to a completely parallel adaptation of 
firing rate.
In Fig.~\ref{positive} we see that adaptation is much stronger for high 
input rather than low input neurons. In this case, $\theta$ at 11.5Hz is a 
fairly low value for neurons to continue to lower their firing rate with 
stepwise adaptation of the chosen ion channels.
This shows the importance of using appropriate 
tuning ('harnessing') mechanisms to make highly nonlinear channels work 
in a purely linear learning context.

Clearly one of the results of learning is an altered spiking behavior
of individual neurons dependent on their history.
It is important to realize that this rule is based on
neural activation, not synaptic input as a learning parameter - since
synaptic input is constant during learning.

\subsection{Positive and Negative Trace Learning}

\begin{figure}[htp]
\begin{center}
\includegraphics[width=0.65\textwidth]{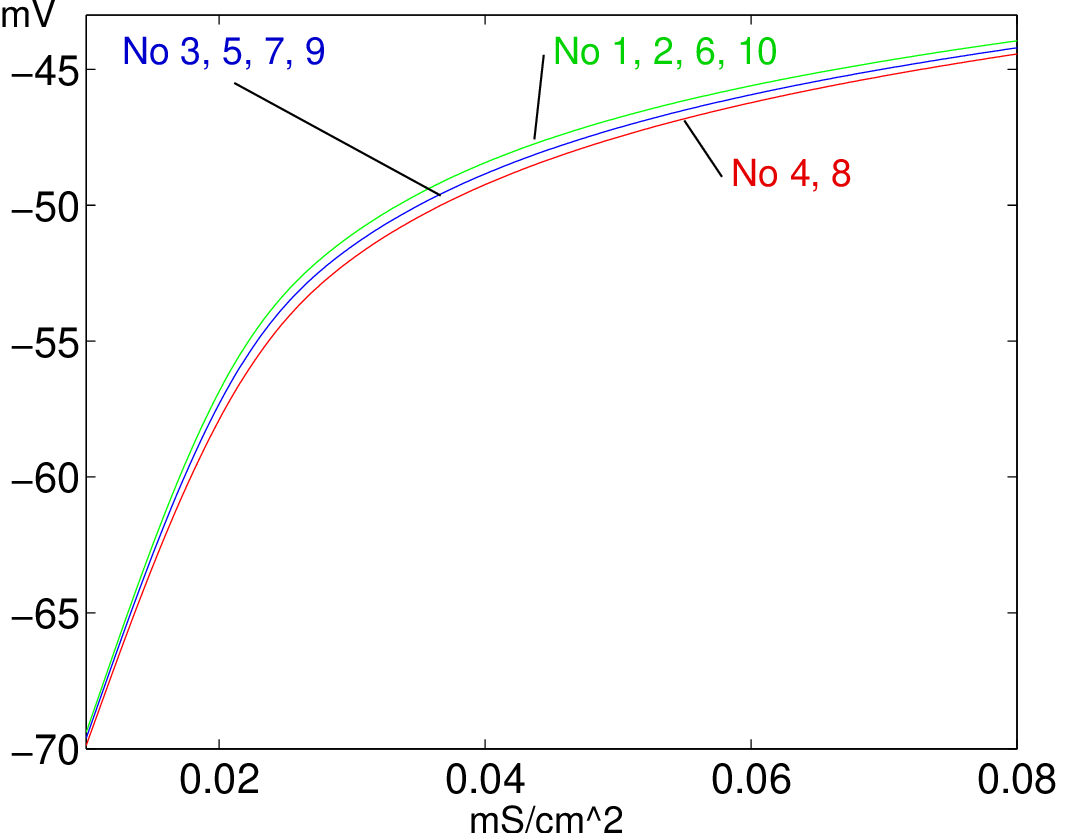}
\end{center}
\caption{
{\bf Negative pattern learning:}
Learning results in different activation functions for high (4, 8), 
medium (3, 5, 7, 9) and low (1, 2, 6, 10) input. 
\label{negative}
}
\end{figure}

We show that this mechanism can be employed not only for positive trace
learning, when excitability adaptation corresponds to frequency
response, but also for negative trace learning, when excitability
adaptation counteracts frequency response and approximates a target 
firing rate $\theta$. This target rate could be set as a result of global 
inhibitory mechanisms corresponding to the expected mean $A_i$ values 
under physiological stimulation. Accordingly, the neuron responds with
decreases
of excitability to high input ranges and increases of excitability to 
low input ranges (Fig.~\ref{negative}).
\begin{equation}
\begin{array}{ll}
\Delta \mu = h(A_n) = - \sigma & \mbox{if}\  A_n > \theta \\
\Delta \mu = h(A_n) = \sigma  &  \mbox{if}\  A_n < \theta\\
\end{array}
\label{eq:An-learning-neg}
\end{equation}

%----------------
% Fig 8
%----------------
%\caption{Negative pattern learning: 
%spike frequency histograms for 10 adaptive neurons ($\theta$=11.5Hz)
%(A) habituation for
%neurons adapted to $P_{learn}$, (B)
%an inverse pattern for $P_{learn}$-adapted neurons tested with $P_{noise}$ 
%and (C) interference (dampening of response) for a new pattern
%$P_{test}$ (naive: line-drawn bars, adaptive: filled bars).
%Average synaptic input ($nA/cm^2$) for each neuron is shown on top. 
%(A) shows a uniform response to patterend synaptic input and (B) a patterned 
%response to uniform (noise) input. (C) shows a difference of response 
%for naive vs. $P_{learn}$-adapted neurons to a new pattern $P_{test}$.
%\label{neg2}

%----------------
% Fig 7
%----------------
%\caption{
%Negative pattern learning: 
%Learning results in different activation functions for high (4,8), 
%medium (3,5,7,9) and low (1,2,6,10) input. 
%\label{negative}
%}

This emphasizes that "homeostatic" responses - adjusting excitability
in the opposite direction to the level of input - can implement trace
learning (pattern learning and feature extraction) as well.

\begin{figure}[htp]
\begin{center}
\includegraphics[width=0.95\textwidth]{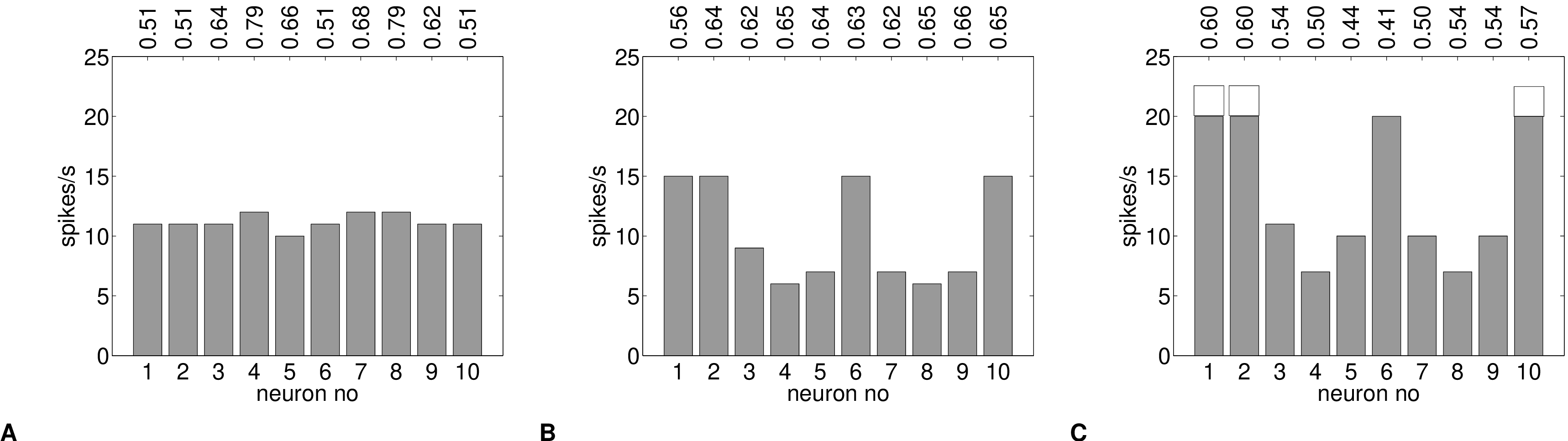} 
\end{center}
\caption{{\bf Negative pattern learning:}
spike frequency histograms for 10 adaptive neurons ($\theta$=11.5Hz)
(A) habituation for
neurons adapted to $P_{learn}$, (B)
an inverse pattern for $P_{learn}$-adapted neurons tested with $P_{noise}$ 
and (C) interference (dampening of response) for a new pattern
$P_{test}$ (naive: line-drawn bars, adaptive: filled bars).
Average synaptic input ($nA/cm^2$) for each neuron is shown on top. 
(A) shows a uniform response to patterend synaptic input and (B) a patterned 
response to uniform (noise) input. (C) shows a difference of response 
for naive vs. $P_{learn}$-adapted neurons to a new pattern $P_{test}$.
\label{neg2}
}
\end{figure}

Negative learning rule results in a mirror image
of parameter values compared to positive learning, 
as shown in Table~\ref{table3}B.
The naive response is the same as before (Fig.~\ref{positive}A).
But here, after adaptation, the neurons have habituated to the input,
and do not produce a strong response anymore 
(Fig.~\ref{neg2}A).  
When neurons are tested with $P_{noise}$, an inverse version of 
the original pattern appears (Fig.~\ref{neg2}B).  
Similarly,
when we apply a different pattern $P_{test}$, we obtain a spike 
histogram, where the learned pattern is overlayed with the new input, 
resulting in a dampening of the frequency response for $P_{test}$
(Fig.~\ref{neg2}C).

For both positive and negative traces, learning is pat\-tern-spe\-ci\-fic, i.e. 
training with homogeneous, fluctuating (high-low) noise, 
such as $P_{noise}$, results in no adjustments (or computes an average).
However, any prolonged sequence of neuron-selective stimulation results in 
neuron-selective patterns. This requires the population to be protected from 
prolonged stimulation with random patterns in a biological setting. 
We may assume most patterns to be meaningful and highly repetitive, while 
the neuron exists in a plastic state, while patterns may be random, when the 
neuron is not plastic (because it is stimulated with highly correlated 
or very low frequency input, saturated in its parameters or undergoes ion 
channel block by selected neuromodulators).

The whole approach to pattern storage and responses elicited to stimulation is 
summarized again in Fig.~\ref{positive} and 
Fig.~\ref{neg2}. We can see that pattern 
storage by changes in intrinsic excitability is useful for a short-term 
buffer system for complete patterns. Patterns are imprinted upon a set of 
neurons and remain available as long as they are not obliterated or overwritten 
by an opposite pattern. Presumably the pattern degrades over time. Training 
with a new pattern - during the period of active maintenance of the pattern - 
would result in cross-activation, i.e. the generation of a mixed pattern. This 
may well be a useful feature of a short-term pattern storage system. It allows 
for pattern integration, or pattern completion from different sources.
Adapting intrinsic excitability has inherent limitations of storage capacity.
We do not fully understand where patterns go after they have passed through 
the intrinsic buffer system, but we assume that synaptic growth, intracellular 
changes and membrane adaptations in a variety of trafficking proteins 
(receptors and channels) all play a role. In the simplest case, the intrinsic 
buffer system serves only to integrate and maintain a pattern of neural 
excitation until all the necessary synaptic adjustments that the memory system 
requires for permanent storage have been made. However, it is not clear, and 
actually highly doubtful at this time that the difference between short-term 
and long-term storage is clear-cut between intrinsic (neuronal) and 
synaptic (esp. glutamatergic synaptic) storage systems.

%-------------
% FIGURE
%-------------
%\includegraphics[width=0.45\textwidth]{FIGURES/fig67_2.eps}
%\caption{Responses to input pattern stimulation (top) after positive learning. First row: naive, untrained neurons. Second row: Neurons trained with the leftmost input pattern. Third row: Neurons trained with the rightmost input pattern. The trained patterns emerge under random stimulation, become stronger after training, and are obliterated by training with the opposite pattern.}
%\label{positive-pattern-learning}

%============================================================
\section{Discussion}
\subsection{Experimental results on induction of intrinsic plasticity}
\label{biolearn}
A number of experimental results show that intrinsic plasticity in MSNs 
may be prominently induced and regulated by intracellular calcium:
It has been shown that e.g. the regulation of delayed rectifier 
$K^+$-channels (Kv2.1 channels) is effectively performed 
by $Ca^{2+}$ influx and 
calcineurin activation in cultured hippocampal neurons,
which can be achieved by glutamate stimulation 
\cite{MisonouHetal2004}. 
The regulation concerns marked
dephosphorylation (reduction of conductance) plus a shift in voltage-dependence.
It has also been shown that 20s of NMDA stimulation, or alternatively, 
increase of intracellular calcium, increases functional dopamine D1 
receptor density at the membrane, which corresponds to an alteration 
in $\kappa$ for D1 parameters, targeting a number of ion channels 
simultaneously \cite{ScottLetal2002}. 
For deep cerebellar neurons, there has recently been some direct evidence on 
the conditions that induce intrinsic plasticity. 
Here, alterations in intrinsic excitability can be induced by bursts of 
EPSPs and IPSPs, accompanied by dendritic calcium transients
\cite{ZhangWetal2004}.
In striatal MSNs, it has been determined that synaptic stimulation at 1Hz 
does not
cause significant calcium signals, but 10Hz stimulation causes moderate
increases, and higher stimulation (up to 100Hz) significantly raises calcium
levels \cite{BonsiPetal2003}. 

In the simulations, neural activation ($A_n$) is estimated from 
the number of spikes generated, measured over the simulated behavior.
In the model case, the membrane potential is not used as a separate parameter,
because membrane potential and spiking behavior are closely linked. 
However, when a neuron exhibits prominent upstates (periods of high membrane 
voltages 
with a variable number of actual spikes), membrane potential may need to 
be treated as an additional, independent component of $A_n$,  
since a great part of the intracellular 
calcium signal in striatal MSNs is being generated from high-voltage activated
NMDA and L-type calcium channels \cite{CarterSabatini2004}. 
The number of spikes produced nonetheless seems important because of the 
phenomenon of backpropagating spikes.
Backpropagating spikes enhance the calcium signal, thus
providing a basis for a prominent role for spiking behavior, or 
firing rate, for defining intracellular calcium.
The presence of backpropagation of spikes has recently been confirmed for 
MSNs \cite{CarterSabatini2004}.

In general, the induction of intrinsic plasticity may 
be linked not only to intracellular calcium.
There exists an intricate intracellular system of interactions between 
diffusible substances like 
calcium and cAMP, as well as a number of crucial proteins (RGS, calcineurin,
PKA, PKC, other kinases and phosphatases) for regulating receptor 
sensitivity and ion channel properties, which are furthermore influenced 
by NM receptor activation. Thus the learning parameter $h$ may 
be analyzed as being dependent not only on 
$A_n$, but also on $[NM]$, and possibly even a third variable for a - 
slowly changing - intracellular state.

\subsection{Synaptic vs. Intrinsic Plasticity}
Learning by intrinsic excitability seems particularly suitable 
for striatal MSNs, since they have few lateral connections, which 
provide only a small part of their total input \cite{TepperJMetal2004}.
When we have strong recurrent 
interaction, as in cortex, intrinsic excitability learning needs to 
adjust
activation functions relative to each other, e.g. to ensure optimal 
distribution of activation functions.
This probably happens in the cortical maps, such as frequency maps in 
auditory cortex \cite{BaoSetal2001}.

In hippocampus, synaptic and intrinsic modulation may potentiate each 
other (E-S potentiation, \cite{ZhangLinden2003}), but 
in other systems (e.g. striatum) 
antagonistic regulation may also exist (such as LTD combined 
with positive learning), 
with effects on the balance of synaptic vs. whole-cell localization for 
the storage of information.

\subsection{Neuromodulation}

When ion channels are regulated by neuromodulation, we can use a factor 
$[NM] \kappa$ - where $[NM]$ is the extracellular concentration of the 
ligand and $\kappa$ the receptor sensitivity (see \ref{variability}, 
Eq.~\ref{eq:mu-kappa-factors}).
$\kappa$ stands for the influence that a NM signal of a certain strength
has on a particular ion channel, i.e. the degree of coupling between
NM receptor ligand binding and ion channel modification \cite{Scheler2004}.
Typically, a signal $[NM]$ will regulate several ion channels in parallel,
but there may be different $\kappa_i$ for each ion channel.

%----------------
% Fig
%----------------

If activation function adaptation proceeds by 
NM-activated $\kappa$ parameters, rather than unconditioned $\mu$ parameters,
response to stimuli will consist of an early, non-modulated component,
where the input pattern is reflected directly in the spiking frequency,
and a later, modulated component, where habituation occurs for a learned 
pattern, or the stored pattern is reflected by
overlaying a new stimulus and the stored pattern.

NM signals orchestrate both adjustments in
activation function and synaptic input,
with NM activation often depressing 
synapses, but increasing the variability in the activation function through 
selected conductance changes (activating $\kappa$-parameters).  As a result, 
the input component of the
signal is reduced in comparison to the stored intrinsic component after NM 
activation.
Presumably, this has a dynamic component, such that for a short time after
a strong signal there is an input-dominant phase which is then followed
by an intrinsic-dominant phase.

\subsection{Homeostasis, Permanence and Information Flow}
There are different ideas at the present time 
what intrinsic plasticity can achieve within a network model of neuronal 
interaction.
%on how synaptic and intrinsic plasticity interact within a network.
Reviews of intrinsic plasticity  \cite{ZhangLinden2003,Xu2005,DaoudalDebanne2003} 
are undecided, whether 
IP acts mainly to maintain homeostasis, adapting to changes in synaptic 
strength by keeping neurons within certain ranges but without significant 
informational capacity, as in the model of \cite{StemmlerKoch99}, or whether
they are themselves capable of being modified in response to particular patterns of activity in ways that facilitate learning and development (\cite{OLeary2011}).
However, as we have shown, homeostatic adaptation does not exclude 
information storage under conditions of conditional read-out.
The synergy between synaptic and intrinsic plasticity may take different 
forms, beyond E-S potentiation. 
Based on experimental evidence in different systems
 \cite{ZhangLinden2003,DestexheMarder2004,Disterhoft2006} have 
listed many possible functions and roles of intrinsic adaptive plasticity,

We have greatly simplified the exposition here by concentrating on spike 
frequency as a major indicator of neural behavior. 
Certainly the type of firing (e.g. burst firing) is also under control 
of neuromodulators, and may be influenced by the distribution and 
density of ion channels.
Single neuron computation is more complex than what can be shown with 
a single compartment model. 
In dendritic computation, the coupling of different compartments may be 
prominently affected by intrinsic plasticity. For instance,
\cite{MisonouHetal2004} showed a 
loss of clustering for $K^+$ channels on the membrane, induced by high glutamate 
stimulation, indicating a possible input-dependent
regulation of dendritic integration.

Studies of concurrent
simulation of synaptic coupling parameters and intrinsic ion channel
conductances has concluded that intrinsic and synaptic 
plasticity can achieve similar effects for network operation 
\cite{PrinzAetal2004}.
We have suggested that
synaptic and intrinsic plasticity can substitute for each other, and
furthermore that this essential functional parallelism could be an indication
for {\it information flow} over time from one modality to the other
\cite{SchelerSocNeurosci}. 
The direction of this information flow may be from intrinsic to synaptic 
for the induction of permanent, morphological changes (such as dendritic spine 
morphology) - however in some systems (e.g. cerebellum) intrinsic 
plasticity may 
also have a permanent component (Purkinje cells) 
\cite{NelsonABetal2003,Coop2010}. 
The detailed interaction between synaptic and 
intrinsic plasticity is still an open question.
Here we have shown a simple, local learning mechanism for intrinsic 
plasticity that allows to store 
pattern information without synaptic plasticity.
This is different from theoretical approaches, where activation functions 
are only being modulated to optimize global measures of information 
transmission between neurons 
while the information is exclusively stored in synaptic weights. 
Further work will be needed to investigate the smooth integration of synaptic 
and intrinsic plasticity and their respective functions in different systems.

%============================================================
\subsection{Conclusions}
We wanted to show quantitatively that IP can have 
significant effects on spike frequency,
dependent on the statistical structure of the input. In particular, low 
correlated input, or input during sensitive (high-voltage membrane) states 
induces the strongest variability of spike responses for different 
activation functions, while 
highly correlated input acts as drivers for neurons, eliminating subtle 
differences in activation function.
We suggested that starting from a very general, natural 
format for a learning rule, which can be biologically motivated, we arrive at 
simple pattern learning, the basis for feature extraction, and realistic 
types of neural behavior: population-wide 
increases/decreases of neural firing rates to novel input stimuli, habituation 
to known stimuli and history-dependent distortions of individual stimuli.
A significant application of this theoretical model exists in the observation 
of pervasive whole-cell adaptations in selected ion channels ($I_{Na}$,
$I_{CaL}$) after cocaine
sensitization \cite{HuXTetal2004,ZhangXFetal2002,ZhangXFetal98}, with 
implications of the type of learning 
that underlies addiction. This would reduce the dynamic range of intrinsic 
plasticity. Potentially, then, learning in striatum is mediated 
in part by intrinsic plasticity (\cite{MahonSetal2003}, and a reduction in inducible 
intrinsic plasticity or dynamic range of intrinsic plasticity after 
cocaine sensitization may contribute 
to the pathology of addiction.

%\clearpage
%%%%%%%%%%%%%%%%%%%%%%
%% The Bibliography %%
%%
\bibliographystyle{spmpsci}

%============================================================
\clearpage
%\section{Tables}
\begin{table*}
\centering
\small
\begin{tabular}{l|lll|llll|llll|l}
\hline\noalign{\smallskip}
I&p&q&$\bar{g}$ &
        $\lambda_\alpha$ & $V_c^\alpha$&$V_i^\alpha$&
                $Eq^{\alpha}$&
        $\lambda_{\beta}$ & $V_c^{\beta}$&$V_i^{\beta}$&$Eq^{\beta}$&
                $E^{rev}$\\[3pt]
\tableheadseprule\noalign{\smallskip}
Na (m)&3& &  35&
              0.1  & 10   &  -28 & 3   &
              4.0  & 18   &  -53 & 1   &
        55 \\

Na (h) & & 1 & &
 0.07 & 20   &  -51 & 1   &
              1    & 10   &  -21 & 2   & \\
%\hline
K & 4 & & 6 &
              0.01 & 10   & -34 & 3    &
              0.125 & 80  & -44 & 1    &
        -90 \\

%\hline
CaL (m) &2& &  0.01&
              0.06  & 3.8   &  -40 & 3   &
              0.94  & 17   &  -88 & 1   &
        140 \\

CaL (h) & & 1 & &
 4.6e-4 & 50   &  -26 & 1   &
              6.5e-3    & 28   &  -28 & 2   & \\
%\hline
leak &   & & 0.04 &
              &   & &    &
              &  &  &    &
        -75 \\
\noalign{\smallskip}\hline
\end{tabular}

\caption{Parameter values for $I_{Na}$, $I_{K}$, $I_{leak}$ as in
%	(Wang and Buzsaki 1996)
\protect{\cite{WangBuzsaki96}}, 
$I_{CaL}$ as in 
%	(Bargas et al.\ 1994, Tsubo et al.\ 2004)
\protect{\cite{BargasJetal94}},
\protect{\cite{TsuboYetal2004}}
}
\label{wangbuzsaki}
\end{table*}

\begin{table*}
\begin{center}
\small
\begin{tabular}{l|lll|lll|lll|lll}
\hline\noalign{\smallskip}

I&p&q&$\bar{g}$&$V_c^{m0}$&$V_i^{m0}$&$Eq^{m0}$&$V_c^{h0}$&$V_i^{h0}$&$Eq^{h0}$&$\tau_m$&$\tau_h$&$E^{rev}$\\[3pt]
\tableheadseprule\noalign{\smallskip}

Kir&1& &0.15&-10&-100&2&    &   & &$<$0.01 &$<$0.01 &-90\\
Af&1&1&0.09&7.5&-33&  2&-7.6&-70&2&1 &25 & -73\\
As&1&1&0.32&13.3&-25.6&2&-10.4&-78.8&2&(a) &(b) &-85\\
Nas&1&&0.11&9.4 &-16.0&2&     &     & &(c)& $<$0.01&40\\

\noalign{\smallskip}\hline
\end{tabular}
\end{center}
\caption{Parameter values for potassium channels $I_{Kir}$, $I_{Af}$, 
$I_{As}$ and a slow sodium 
channel $I_{Nas}$   
%	cf.~(Mahon et al.\ 2000b, Gruber et al.\ 2003),
cf.\ \cite{MahonSetal2000,GruberAJetal2003}, 
where \protect{\small
(a) $\tau_m = 131.4/(\exp(-(V_m + 37.4)/27.3)  + \exp((V_m + 37.4)/27.3)) $ (b) $\tau_h = 179.0 + 293.0*exp(-((V_m + 38.2)/28)^2) * ((V_m + 38.2)/28)$ (c) $\tau_m = 637.8/(\exp(-(V_m + 33.5)/26.3)  + \exp((V_m + 33.5)/26.3)) $}}
\label{Mahon}
\end{table*}

\begin{table}
\small
\begin{minipage}[t]{0.40\textwidth}
%\begin{tabular}{lp{6em}p{6em}p{6em}}
\begin{tabular}{lp{3em}p{3em}p{3em}}
\hline\noalign{\smallskip}
no& $\mu_{CaL}$ & $\mu_{As}$ & $\mu_{Na}$ \\[3pt]
\tableheadseprule\noalign{\smallskip}
1 & 0.9/0.8 & 1.2/1.4 & 0.9/0.8 \\ %L
2 & 0.9/0.8 & 1.2/1.4 & 0.9/0.8 \\ %L
3 & 1.0 & 1.0   & 1.0 \\ %M
4 & 1.1/1.2 & 0.8/0.6   & 1.1/1.2 \\ %H
5 & 1.0 & 1.0   & 1.0 \\ %M
6 & 0.9/0.8 & 1.2/1.4 & 0.9/0.8 \\ %L
7 & 1.0 & 1.0   & 1.0 \\ %M
8 & 0.9/0.8 & 1.2/1.4 & 0.9/0.8 \\ %L
9 & 1.0 & 1.0   & 1.0 \\ %M
10 & 0.9/0.8 & 1.2/1.4 & 0.9/0.8 \\ %L
\noalign{\smallskip}\hline
\end{tabular}\\[1.5ex]
\noindent A
\end{minipage}
%
%\ \\[2.5ex]
%
\hfill
\begin{minipage}[t]{0.40\textwidth}
%\begin{tabular}{lp{6em}p{6em}p{6em}}
\begin{tabular}{lp{3em}p{3em}p{3em}}
\hline\noalign{\smallskip}
no& $\mu_{CaL}$ & $\mu_{As}$ & $\mu_{Na}$ \\[3pt]
\tableheadseprule\noalign{\smallskip}

1 & 1.4 & 0.7   & 1.0 \\ %L
2 & 1.4 & 0.7   & 1.1 \\ %L
3 & 0.9 & 1.1   & 1.0 \\ %M
4 & 0.6 & 1.4   & 0.8 \\ %H
5 & 1.1 & 0.8   & 0.9 \\ %M
6 & 1.4 & 0.7   & 1.1 \\ %L
7 & 1.1 & 0.8   & 0.9 \\ %M
8 & 0.7 & 1.4   & 0.8 \\ %H
9 & 1.0 & 0.9   & 0.9 \\ %M
10& 1.4 & 0.7   & 1.1 \\ %L
\noalign{\smallskip}\hline
\end{tabular}\\[1.5ex]
\noindent B
\end{minipage}
\caption{Pattern learning: 
Parameter values (A) for positive learning with (weak/strong) adaptation 
of $\mu$ values and (B) negative learning. CaL and Na channels are adapted 
in the opposite direction to K channels.}
\label{table3}
\label{table4}
\end{table}

%============================================================

%--------------------------------------------------------------

\end{document}